# Higher-order topological and nodal superconducting transition-metal sulfides MS (M = Nb and Ta)

Yipeng An,[1,*] Juncai Chen,[1] Yong Yan,[1] Jinfeng Wang,[1] Yinong Zhou,[2] Zhengxuan Wang,[1] Chunlan Ma,[3] Tianxing Wang,[1] Ruqian Wu,[2,†] and Wuming Liu[4,‡]

[1]*School of Physics, Henan Normal University, Xinxiang 453007, China*

[2]*Department of Physics and Astronomy, University of California, Irvine 92697, USA*

[3]*School of Physics and Technology, Suzhou University of Science and Technology, Suzhou 215009, China*

[4]*Beijing National Laboratory for Condensed Matter Physics, Institute of Physics, Chinese Academy of Sciences, Beijing 100190, China*

Intrinsic topological superconducting materials are exotic and vital to develop the next-generation topological superconducting devices, topological quantum calculations, and quantum information technologies. Here, we predict the topological and nodal superconductivity of NiAs-type MS (M = Nb and Ta) transition-metal sulfides. We reveal their higher-order topology nature with an index of $Z_4 = 2$. These materials have a higher $T_c$ than the Nb or Ta metal superconductors due to their flat-band and strong electron-phonon coupling nature. Electron doping and lighter isotopes can effectively enhance the $T_c$. Our findings show that the MS (M = Nb and Ta) systems can be new platforms to study exotic physics in the higher-order topological superconductors, and provide a theoretical support to utilize them as the topological superconducting devices in the field of advanced topological quantum calculations and information technologies.

## I. INTRODUCTION

Topological materials have aroused great interest in recent years for the realization of integer quantum Hall effects and dissipationless charge and spin transport. Various exotic topological materials have been synthesized or predicted, such as topological insulators [1-3], topological semimetals [4-7], topological superconductors [8-13], topological excitations [14,15], topological phononic crystals [16-19], topological photonic crystals [20-22], and topological time crystals [23-25]. Particularly, it is significant to advance the research on topological superconductors, whose bulk has the superconducting state and surface has a topology-protected gapless metallic state. They not only have zero-resistance in both bulk and surface [26], excellent for saving energy, but they also are imperative for realizing the Majorana zero modes [27] and developing the advanced topological superconducting devices and topological quantum circuits for quantum information technologies[28]. Furthermore, nodal superconductors, as a new class of topological superconductors, are more common than the fully gapped superconductors [29]. The nodal topological superconductors exhibit rich nontrivial topological characteristics up to their node type (i.e., point, line, or surface), such as the nodal noncentrosymmetric superconductors [30,31], high-$T_c$ $d$-wave superconductors [32,33], Weyl superconductors [34,35], and heavy Fermion systems [36,37]. The nontrivial topology of nodal superconductors can often be observed at the surface in the form of Majorana flat-band states, arc surface states, or cone states [30].

There are two main approaches to realize the topological superconducting phase. The first is based on the proximity effect. One can merge a conventional superconductor with a topological insulator into an artificial heterostructure that allows the coexistence of the superconducting and topological states [9,38]. The second is the intrinsic topological superconductor, which can be obtained by properly doping the topologically-trivial superconductors or topological insulators [39]. Nevertheless, the observation of either intrinsic or artificial topological superconducting materials has been rather rare.

* ypan@htu.edu.cn
† wur@uci.edu
‡ wliu@iphy.ac.cn

Niobium (Nb), a prototypical superconductor for the development of superconducting device, has the highest transition temperature ($T_c$ = 9.2 K) in all elemental superconducting metals [40]. Various methods have been attempted to enhance its $T_c$, such as doping metal impurities [41], synthesizing Nb-based alloys and compounds [42-45], as well as applying high pressure [46,47] and magnetic field [48]. Again, making significant improvement or incorporating topological features in Nb-based superconductors has not been very successful.

In this paper, we discover the higher-order topological and nodal superconductivity of non-van der Waals layered structure NiAs-type NbS by means of *ab initio* calculations. NbS has a higher $T_c$ (16.27 K) than Nb metal ($T_c$ = 9.2 K) superconductor under normal pressure, due to its strong electron-phonon coupling (EPC) nature. In addition, we predict that NiAs-type TaS is also a higher-order topological and nodal superconductor with $T_c$ = 12.30 K. Electron doping and lighter isotopes can effectively enhance their $T_c$, while the external pressure not. These results shed light on the mechanism of generating topological superconducting states in layered materials and open a vista for the search of new families of topological superconductors.

## II. METHODS

The *ab initio* calculations are carried out with the QUANTUM ESPRESSO (QE) code [49] unless stated otherwise. The Perdew-Burke-Ernzerhof function of the generalized gradient approximation [50-54] is adopted for the MS (M = Nb and Ta) compounds. The SG15 optimized norm-conserving Vanderbilt pseudopotentials [55-57] are used to describe the effect of core electrons. The kinetic energy cutoff for wavefunctions (charge density and potential) is set to 80 (320) Ry. A 12 × 12 × 8 Monkhorst-Pack *k*-point grid is used for the self-consistent calculations, and a denser grid (24 × 24 × 16) is adopted to calculate the density of states (DOS) and bands, while coarse *q*-point grid (6 × 6 × 4) for the phonon calculations to balance the calculation accuracy and cost. The total energy tolerance and residual force on each atom are < $10^{-10}$ Ry and $10^{-8}$ Ry Bohr$^{-1}$ in the structure relaxation. The phonon properties are obtained by the density functional perturbation theory [58]. The optimized tetrahedron method [59] is employed to integrate the Brillouin zone and get the EPC interaction.

The surface states are obtained by using the WANNIER90 [60] and WANNIERTOOLS [61] codes including the spin-orbit coupling (SOC) effect. Higher-order topology properties are obtained by using the QEIRREPS code [62] based on the self-consistent results of QE code [49]. The phonon-mediated superconductivity calculations are performed by the density functional theory for superconductors (SCDFT) method as implemented in the SUPERCONDUCTING-TOOLKIT code [63-69]. Note the phonon-related calculations are performed without including the SOC effect due to it remains the band structure unchanged [see Fig. S1(a) of the Supplemental Material [70] for the result of NbS with weak SOC effect] and is less important in describing the vibrational and superconductivity properties [71]. In addition, the present work does not include the Hubbard U parameter, since it (U = 3 eV on the Nb-*d* states) slightly underestimates the lattice constant and gives the consistent band structure near the $E_F$ [see Fig. S1(b) of the Supplemental Material [70]] which quantitatively retains the conclusions. To study the isotope effect on the superconducting $T_c$, it is easily achieved by specifying different atomic mass for the different isotopes in the phonon calculations. More calculation details can be found in the Supplemental Material [70].

## III. RESULTS AND DISCUSSION

### A. Electronic structures of NbS bulk

Figure 1(a) shows the hexagonal NiAs-type NbS bulk structure, which has the space-inversion symmetry with the space group of *P*6$_3$/*mmc* (No. 194). Our calculated lattice constants are *a* = 3.29 Å and *c* = 6.67 Å, both in consistent with the experimental data [72]. The most remarkable feature in the electronic structure is that NbS has a flat-band (FB) near the Fermi level ($E_F$) in the path of *Γ-M-K* [see band 19 in Fig. 1(b)], which gives rise to the large electronic density of states (DOS) near the $E_F$ and plays a key role in the superconductivity [74,75]. Nb atoms (i.e., Nb$_{4d}$ orbitals) give the main contribution to the FB and electronic states near $E_F$ according to the element-projected band structure and DOS [see Fig. 1(b)].

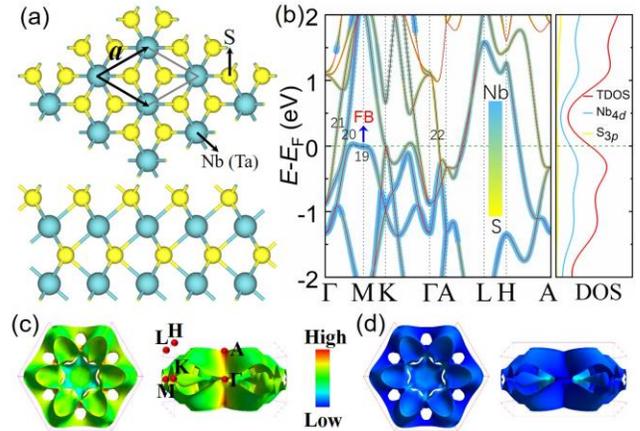

FIG. 1. (a) Top and side views of MS (M = Nb and Ta) bulk. (b) Projected band and density of states for the Nb$_{4d}$ and S$_{3p}$ orbitals of NbS, as well as total density of states (TDOS). 19-22 refer to the four bands crossing $E_F$. Top and side views of Fermi velocity |$v_F$| projected on Fermi surfaces drawn by using the FERMISURFER code [73] for (c) Nb$_{4d}$ and (d) S$_{3p}$ orbitals of NbS. High symmetry points of the first Brillouin zone of NbS is shown in (c).

Moreover, the Fermi velocity $|v_F|$ of Nb$_{4d}$ orbitals is larger than that of S$_{3p}$ orbitals from their projections on the lotus-like Fermi surface [see Figs. 1(c) and 1(d)] formed by the four bands crossing the $E_F$ [see bands 19 to 22 in Fig. 1(b)]. Both of them display a large anisotropy in the first Brillion zone such as, the ratio between the maximum to the minimum velocity for the Nb$_{4d}$ orbitals is 51. Their projections on the four bands are supplied in the Fig. S2 of Supplemental Material [70]. These demonstrate that Nb$_{4d}$ orbitals dominant the carrier motion and electroconductibility of NbS. Note that these three steep bands crossing $E_F$ [see bands 20 to 22 in Fig. 1(b)] play a major role in the high Fermi velocity. Such coexistence of flat and steep bands with localized and ultramobile electrons has been shown to enhance superconductivity [76-79].

### B. Phonons and electron-phonon coupling of NbS bulk

The FB at $E_F$ of NbS arouses our attention about the possibility of having phonon-mediated superconductivity via the EPC. The flat bands with localized electrons and steep bands with high Fermi velocity interact differently with phonons [80]. The coexistence of flat and steep bands near the $E_F$ is important for the "flat-band/steep-band" scenario of superconductivity [76-81]. Therefore, in the hexagonal NiAs-type NbS bulk structure, the flat band in the path of $\Gamma$-$M$-$K$ and the steep bands with the high Fermi velocity could lead to a higher $T_c$ with the unconventional EPC. The phonon dispersion of NbS bulk is first examined, which is completely positive under static zero-pressure [Fig. 2(a)], confirming its dynamic stability. All the three acoustic branches and the three optical branches with lowest frequency are mostly contributed from the Nb atoms, and the left six high-frequency optical branches mainly stem from the vibrations of light S atoms, based on their projected phonon DOS (PHDOS). The out-of-plane acoustic (ZA) mode (i.e., along the $\Gamma$-$M$ and $A$-$L$ paths) has strong EPC interaction $\lambda_{qv}$, which can give the main contribution to the phonon-mediated superconductivity of NbS. In addition, the high-frequency optical branches dominated by S atom vibrations have larger linewidths $\gamma_{qv}$ (proportional to inverse phonon lifetime) than low-frequency optical branches and all acoustic branches contributed by Nb vibrations. It is reasonable that the phonons with higher frequency contributed by the lighter atoms often have shorter lifetime (larger linewidths) and are easier to be scattered.

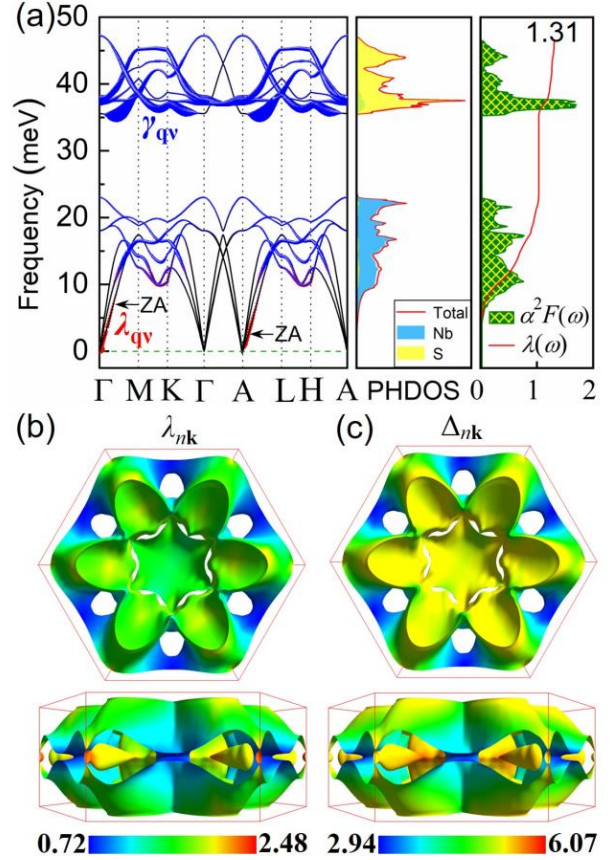

FIG. 2. (a) Phonon band with electron-phonon coupling $\lambda_{qv}$ (red balls) and phonon linewidth $\gamma_{qv}$ (blue balls, by a factor of 0.4 with unit of meV), projected phonon density of states (PHDOS), and Frequency-dependent Eliashberg spectral functions $\alpha^2F(\omega)$ with the cumulative EPC strength $\lambda(\omega)$ of NbS. Top and side views of EPC $\lambda_{nk}$ (b) and Superconducting gap $\Delta_{nk}$ (c) on Fermi surfaces.

We now turn to the EPC of NbS. The calculated frequency-dependent Eliashberg electron-phonon spectral function $\alpha^2F(\omega)$ and the cumulative EPC strength $\lambda(\omega)$ are shown in the right panel of Fig. 2(a). The Nb vibrations make the main contributions to the EPC. They lead to the first significant increase of the cumulative EPC strength ($\lambda$ = 1.03) and account for 79% of the total EPC strength of $\lambda$ = 1.31. The high-frequency optical branches from the S vibrations induce the second step of EPC, with an average phonon frequency $\omega_{ln}$ = 16 meV (184 K), which is proportional to the $T_c$ based on the McMillan's formula and defined as

$$\omega_{ln} = \exp\left[\frac{1}{\lambda}\sum_{qv}\ln(\omega_{qv})\lambda_{qv}\right]. \quad (1)$$

In detail, the EPC has a strong anisotropy [see Fig. 2(b)] in the first Brillion zone based on the SCDFT method. The ratio between the maximum to the minimum $\lambda_{nk}$ (band index $n$ and wave number **k** dependent) projected on the Fermi surfaces is 3.4. Both the flat band 19 and the steep bands 20 to 22 give the significant contributions to the EPC, whose projections on the four bands crossing $E_F$ can be found in the Fig. S3 of Supplemental Material [70]. The average value ($\lambda_{avg}$) of EPC is 1.38 (> 1.31 and 1), demonstrating its strong and anisotropic EPC superconductivity.

## C. Superconductivities and higher-order topology properties of NbS bulk

In order to evaluate the EPC-induced superconductivity of NbS, we examine its superconducting gap by using the SCDFT method [63-69]. The anisotropic EPC nature of NbS leads to the anisotropic superconducting gap $\Delta_{n\mathbf{k}}$ (projected on the Fermi surfaces) [Fig. 2(c)], whose ratio between the maximum to the minimum value is 2.1. The average gap value $\Delta_{avg}$ is 4.44 meV. The $\Delta_{n\mathbf{k}}$ of NbS has a full gap $s$-wave feature, and its projections on the four bands crossing $E_F$ can be found in the Fig. S4 of Supplemental Material [70]. Using the SCDFT and bisection method [63-69], the calculated intrinsic superconducting transition temperature $T_c$ of NbS is 16.27 K, which is significantly enhanced compared with Nb superconductor (9.2 K), suggesting the promising application foreground of NbS as a superconductor.

Next, we move to the topology properties of NbS. Interestingly, NbS shows a higher-order topology nature with the index of $Z_4 = 2$, which is defined as the sum of the inversion parities of all occupied bands at all eight points with the time-reversal invariant momenta, as obtained by using the QEIRREPS code [62] combing with the spin-orbital coupling (SOC) results from the QE code [49,82,83].

We further calculate its surface states and Fermi arcs (at the $E_F$) projected on the (001) surface [see Figs. 3(a) and 3(b)], which are obtained by WANNIER90 [60] and WANNIERTOOLS [61] codes including SOC effect. One can see clearly that some surface states (red curves) of the (001) surface cross the $E_F$ [Fig. 3(a)]. A few closed Fermi arcs are observed and exhibit the hexagonal shape [Fig. 3(b)]. We also show the surface states on the (100) plane as shown in the Fig. 3(c). One can see that some flat-band surface states appear around the $E_F$, which could play a key role in the superconductivity of NbS slab. A few open and closed arc surface states are also observed in the Fermi arcs map [Fig. 3(d)].

## D. Topological and nodal superconducting properties of NbS bulk

As the superconductivity and higher-order topology coexist in NbS, we are inspired to further examine its superconductor nature by using symmetry indicators (SIs) [84-86] such as, topological superconductor, nodal superconductor, or only normal superconductor. Because the former is a crucial material to prepare quantum devices utilized in the quantum computation and quantum information. By examining irreducible representations of space groups, the SIs method can efficiently determine the topological properties of materials and has been employed in various materials including superconductors [87-91]. The symmetry-based diagnosis of topological and nodal superconductor nature for NbS including the SOC effect is obtained by using the QEIRREPS [62] and TOPOLOGICAL SUPERCON [92] tools (see Table I).

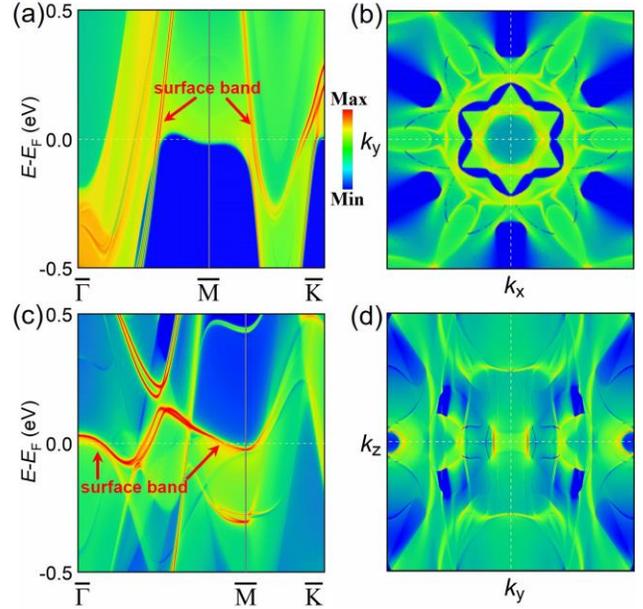

FIG. 3. (a) and (c) are the surface states, and (b) and (d) are Fermi arcs (open and closed red arcs) at $E_F$ projected on semi-infinite (001) and (100) surfaces of NbS, respectively.

Generally, a superconductor (for each pairing symmetry) falls into one of the following four categories: (I) representation-enforced nodal superconductor (NSC); (II) topological NSC (TNSC) or symmetry-diagnosable topological superconductor (TSC); (III) topology-trivial or not symmetry-diagnosable TSC; (IV) silent for the trivial pairing (due to no band labels can be defined at any high-symmetry momenta) [91].

Hexagonal NbS has the $D_{6h}$ point group and contains eight 1D single-valued representations (see Table I), all of which preserve time-reversal symmetry. It is mostly a representation-enforced NSC (i.e., case I) except that the pairing is $A_{1u}$ and $A_{1g}$. The superconducting phases of NbS are predicted to have node lines ($L$) for *even*-parity and node points ($P$) for *odd*-parity cases, respectively. Compatibility relations (CRs, or called symmetry constrains) along various lines are broken for the first six pairing symmetry. On the contrary, all CRs are satisfied only for the $A_{1u}$ pairing. Therefore, one can diagnose the topology by the method of SIs. A fully gapped TSC phase (i.e., case II) is observed for the $A_{1u}$ pairing that belongs to the entry (0, 0, 1, 1, 0, 8). Last, the trivial $A_{1g}$ pairing symmetry is part of case IV for which no band labels can be well defined [91]. These results demonstrate the nodal and topological superconductor nature of NbS, which can be utilized for developing the advanced topological superconducting devices and quantum circuits for quantum information technologies [28] through its topologically protected exotic surface states.

TABLE I. Results of symmetry-based diagnosis for higher-order topological and nodal superconductor NbS. The first column is the point group of NbS. The second and third column lists its pairing symmetries and corresponding cases. Symbols [P] and [L] for case I indicate the shape of nodes, point and line, respectively. The last column refers to the paths where compatibility relations (CRs, or called symmetry constrains) are violated for case I and the entry of symmetry indicators for case II.

| PG | Pairing | Case | Nodes | Topology |
|---|---|---|---|---|
| | $B_{1u}$ | I [P] | Γ-A | NSC |
| | $B_{1g}$ | I [L] | Γ-A, M-L | NSC |
| | $B_{2u}$ | I [P] | Γ-A | NSC |
| $D_{6h}$ | $B_{2g}$ | I [L] | Γ-A, M-L, K-H | NSC |
| | $A_{2u}$ | I [P] | Γ-A, M-L | NSC |
| | $A_{2g}$ | I [L] | A-H, Γ-A, M-L, K-H | NSC |
| | $A_{1u}$ | II | (0, 0, 1, 1, 0, 8) | TNSC or TSC |
| | $A_{1g}$ | IV | … | … |

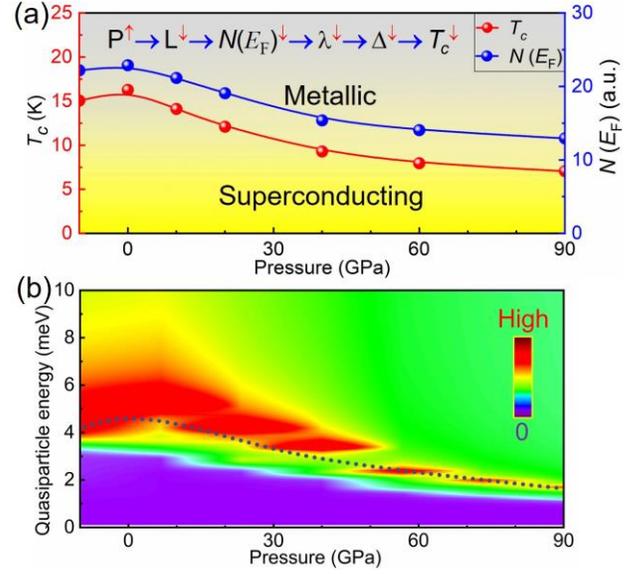

FIG. 4. (a) Pressure-dependent $T_c$ and $N(E_F)$, as well as the phase diagram of superconducting and metallic states. The embedded illustration from "P" (i.e., pressure) to "$T_c$" indicates the control mechanism of superconductivity by external pressures. (b) Pressure-dependent normalized superconducting quasiparticle density of states (QPDOS) at 0.1 K (equivalent to tunnel conductance spectrum observed in experiments). The black dashed line depicts the trend of single-gap peak vs pressure.

### E. Quantum control of the superconductivities of NbS bulk

We further try to increase the $T_c$ of NbS by using the three approaches respectively: applying pressure, carrier doping, and isotope replacement. Generally, the $T_c$ of conventional superconductors increases with the $N(E_F)$ (i.e., DOS at $E_F$ in the superconducting state). The pressure may slightly decrease the $T_c$ of NbS due to the depressed $N(E_F)$ [Fig. 4(a)] which is proportional to EPC $\lambda$. The $N(E_F)$ is sensitive to the external pressure, as the lattice constants and average superconducting gap $\Delta_{avg}$ decrease monotonously and FB gradually shifts down when the external hydrostatic pressure increases (see Figs. S5 and S6 of Supplemental Material [70]). Namely, the increasing pressures lead to the monotonous decrease of lattice (L), $N(E_F)$, EPC ($\lambda$), superconducting gap $\Delta$, and $T_c$. Note the negative pressure also decreases its $T_c$ and can induce an unstable phase when the pressure is beyond −10 GPa, because imaginary frequencies appear such as at −20 GPa (see the Fig. S7 of Supplemental Material [70]).

One dominant approach to probe the superconductors (such as their surface states) is the conductance spectroscopy experiment. Because the conductance is sensitive to the surface DOS of superconductors which can provide direct evidence of surface states. We plot the calculated pressure-dependent normalized superconducting quasiparticle DOS (QPDOS) at 0.1 K in Fig. 4(b) that can be devoted to comparing directly with the experimental tunneling conductance. The QPDOS shows a clear gap feature (purple region) which often arises from the resonant tunneling through the surface states of a superconductor. The cockscomb ridge map that is comprised by the QPDOS peaks suggests the single-gap nature of NbS superconductor, which is induced by its Nb$_{4d}$ orbitals and equivalent to its superconducting gap. The QPDOS peak becomes narrower and lower with the increase of pressure, showing that the inadequate pressure fails to enhance its $T_c$.

We next try to enhance the $T_c$ of NbS by increasing the $N(E_F)$ via carrier doping so as to shift the $E_F$. Here we use a jellium model to simulate the electron and hole doping effect by directly reducing and increasing the total electron number of the system (see Table II). The electron and phonon band structures under the various carrier doping concentrations are shown in the Fig. S8 of Supplemental

TABLE II. Superconducting properties of NbS under various hole and electron doping concentrations, including the data of $N(E_F)$ (in states/eV), EPC $\lambda$, superconducting gap $\Delta$ (in meV), and $T_c$ (in K).

| Doping | $N(E_F)$ | $\lambda$ | $\Delta$ | $T_c$ |
|---|---|---|---|---|
| 0.04 $h$/cell | 22.43 | 1.21 | 3.97 | 15.31 |
| 0.02 $h$/cell | 22.78 | 1.29 | 4.19 | 15.79 |
| 0.00 | 22.87 | 1.38 | 4.44 | 16.27 |
| 0.02 $e$/cell | 23.26 | 1.54 | 4.65 | 16.57 |
| 0.04 $e$/cell | 23.60 | 1.56 | 4.75 | 16.92 |
| 0.05 $e$/cell | 23.76 | 1.67 | 4.90 | 17.18 |
| 0.06 $e$/cell | 23.93 | 1.84 | 5.06 | 17.55 |

Material [70]. The phonon dispersion curves are robust and slightly becomes softened and hardens for the electron and hole doping respectively, demonstrating its dynamic stability under the carrier doping. The electron and hole doping shift the $E_F$ upward and downward with respect to the intrinsic system. As expected, the $N(E_F)$ increases gradually with the increase of electron doping concentration. In contrast, it decreases monotonously when the hole doping concentration is enlarged. As a result, the superconductive parameters of EPC $\lambda$, superconducting gap $\Delta$, and $T_c$ exhibit the same increase (decrease) trend with the increase of electron (hole) doping concentrations. For instance, the $T_c$ can be enhanced to 17.55 K under a high electron doping concentration of 0.06 $e$/cell. This demonstrates that only the electron doping can effectively enhance the $T_c$ of NbS.

The phonon-mediated superconductivity is related to the whole atomic mass including its neutron number. Here we turn our attention to the isotope effect to the $T_c$ of NbS. We focus on the isotopes of Nb with the mass from 89 to 96, each of which has a long half-life. The results demonstrate that the lighter NbS isotopes have higher $T_c$, vice versa. For instance, the $T_c$ of $^{89}$NbS increases to 16.53 K, while it decreases to 16.04 K for $^{96}$NbS.

### F. Higher-order topological and nodal superconductivities of TaS bulk

Finally, we also further study the topological superconductivities of NiAs-type TaS bulk as a MS-family transition-metal sulfide. It has a similar electronic structure but with a slightly curved flat-band around the $M$ point near the $E_F$, which is dominant by the Ta$_{5d}$ orbitals that also give the most contributions to the Fermi velocity (see the Figs. S9 and S10 of Supplemental Material [70]). TaS also shows a topologically nontrivial nature with a higher-order index of $Z_4 = 2$, including the SOC effect (see the Fig. S11 of Supplemental Material [70]), whose surface states and Fermi arcs projected on the semi-infinite (001) and (100) surfaces are shown in the Fig. S12 of Supplemental Material [70]. Compared with NbS, the curved surface bands observed on the (100) surface around $\Gamma$ point demonstrate its depressed superconductivity. In addition, TaS has a depressed $N(E_F)$ due to the appearance of curved flat-band, although it has a larger EPC strength ($\lambda = 1.45$). As a result, the $T_c$ of TaS is reduced to 12.30 K smaller than NbS but larger than Ta metal superconductor (4.5 K) (see the Figs. S13 to S17 of Supplemental Material [70]). It has the similar topological and nodal superconductivities to NbS but with different paths where compatibility relations are violated for case I and entry of symmetry indicators for case II (see the Table S1 of Supplemental Material [70]).

## IV. CONCLUSIONS

In conclusion, we find the topological and nodal superconductors based on the non-van der Waals layered MS (M = Nb and Ta) materials. Their superconductivity and higher-order nontrivial topology properties are revealed by means of *ab initio* methods, and thoroughly analyzed from the aspects of symmetry and electron-phonon coupling. They are single-gap $s$-wave superconductors with higher $T_c$ than Nb and Ta metals due to the strong electron-phonon coupling feature. The electron doping and lighter isotopes can effectively enhance their $T_c$, while the external pressure is nevertheless ineffective to further enhance their $T_c$. Our results demonstrate that the MS (M = Nb and Ta) systems can become new platforms to develop quantum materials and devices based on their unique higher-order topological and nodal superconducting properties, and pave a way to further study the topological and superconducting properties of NiAs-type transition-metal chalcogenides.

## ACKNOWLEDGMENTS

Work in China was supported by the National Natural Science Foundation of China (Grant Nos. 12274117, 61835013, 12174461, and 12234012), the National Key R&D Program of China (Grant Nos. 2021YFA1400900, 2021YFA0718300, 2021YFA1402100), the Natural Science Foundation of Henan Province (Grant No. 202300410226), the Young Top-notch Talents Project of Henan Province (2021 year), Space Application System of China Manned Space Program, Henan Center for Outstanding Overseas Scientists (No. GZS2023007), and the HPCC of HNU. Work at UCI was supported by the U.S. DOE, Office of Science (Grant No. DE-FG02-05ER46237). We thank M. Kawamura and S. Ono at the University of Tokyo, B. B. Ruan at Institute of Physics of CAS for helpful discussions.